\documentclass[9pt,conference]{IEEEtran}

\usepackage{waspaa25} 

\usepackage{bm} 
\usepackage{nicefrac}
\usepackage{pifont}
\def\BibTeX{{\rm B\kern-.05em{\sc i\kern-.025em b}\kern-.08em
    T\kern-.1667em\lower.7ex\hbox{E}\kern-.125emX}}
\usepackage{pifont}
\newcommand{\xmark}{\ding{55}}%

\newcommand{\fix}[1]{\textcolor{black}{#1}}

\title{Generating Separated Singing Vocals Using a Diffusion Model Conditioned on Music Mixtures}

\name{Genís Plaja-Roglans$^{1,2}$,
      Yun-Ning Hung$^{1}$,
      Xavier Serra$^{2}$,
      Igor Pereira$^{1}$}
\address{$^{1}$Music.AI, Salt Lake City, USA
\\$^{2}$Music Technology Group, Universitat Pompeu Fabra, Spain}

\begin{document}

\maketitle

\begin{abstract}
Separating the individual elements in a musical mixture is an essential process for music analysis and practice.
%
%
%
While this is generally addressed using neural networks optimized to mask or transform the time-frequency representation of a mixture to extract the target sources,
the flexibility and generalization capabilities of generative diffusion models \fix{are giving rise to} a novel class of solutions for this complicated task.
In this work, we explore singing voice separation from real music recordings using a diffusion model which is trained to generate the solo vocals conditioned on the corresponding mixture. 
Our approach improves upon prior generative systems and achieves competitive objective scores against non-generative baselines when trained with supplementary data. The iterative nature of diffusion sampling enables the user to control the quality-efficiency trade-off, and also refine the output when needed. We present an ablation study of the sampling algorithm, highlighting the effects of the user-configurable parameters. 

\end{abstract}

\section{Introduction}\label{sec:intro}
Music source separation (MSS) is the task of estimating the individual elements in a music mixture~\cite{unet}, 
a crucial process for music creation and analysis~\cite{plaja2023carnatic}.
MSS is typically addressed using neural networks that mask or transform the time-frequency representation of a mixture to extract individual sources~\cite{sdx_2023}.
%
%
%
However, this may be limited by the prominent source overlap in music, and further refinement is normally beneficial~\cite{spleeter,d3net,msg,lutati2024separate}. 
\fix{Moreover, handling phase adds an extra step that is prone to introducing artifacts~\cite{jansson-complex}.}
Finally, these systems rely on access to \fix{a complete set of} isolated source recordings that linearly sum to the mixture, however, this is often difficult to collect~\cite{instglow}.

Interestingly, deep generative modeling has advanced notably with the advent of denoising diffusion probabilistic models (DDPM)~\cite{diff_beat_gans}. 
%
%
The iterative sampling scheme of DDPM has shown capability to address audio inverse problems~\cite{moliner_inverse, universe_serra}, has been leveraged for refining predictive MSS systems~\cite{plaja-inspired}, and has recently outperformed prior generative approaches for this problem~\cite{pritish_sep, gan_prior, instglow, postolache2023latent}. In fact, DDPM are considered a promising technology for MSS given its flexibility and generalization~\cite{araki20253030years}.
%
%
%
%
However, existing DDPM for separation primarily focus on synthetic instrumental music 
and have room for improvement on efficiency and performance~\cite{mariani2023multi,msdm_ldm,plaja-ldm}.

In this work, we explore again the potential of DDPM to advance towards competitive generative-based MSS. Specifically, we address the following goals:
\textbf{(1)} separating the singing voice–-a complex but important source–-from real music mixtures, while solely requiring \fix{corresponding} pairs of vocals and mixtures for training, \textbf{(2)} developing a lighter model that can sample satisfactory generations in fewer sampling steps than prior work\fix{~\cite{gan_prior, instglow, mariani2023multi}}, and \textbf{(3)} exploring a flexible sampling process to \fix{refine the fidelity of} the generated vocals without further training or any additional post-processing system. 
The proposed approach is based on a generator network that is trained to sample solo singing voice signals,
\fix{guided by a coupled conditioning module that injects music mixtures into the generator–-steering the generation toward vocals consistent with the mixture.}
We employ our own tailored version of the conditioning module in \cite{universe_serra}.

The trained system is able to generate acceptable separations in $\approx20$ diffusion sampling steps. However, we improve the generated vocals relying on an ablation study on the sampling algorithm, which includes modifications to refine the separations in a chosen frequency range, with no need for an additional system to process the outputs.

Our system outperforms existing generative baselines for vocal separation on open multi-stem datasets. If trained with supplementary data, we achieve competitive scores for objective separation metrics against the non-generative literature–-despite potential inherent penalization~\cite{gan_prior}–-showing remarkable cleanliness and correspondence to the reference mixture. 
We make our code available in a modular fashion to build our system but also to perform further research, together with further resources.\footnote{\url{https://github.com/weAreMusicAI/dmx-diffusion}}

\begin{figure*}
 \centerline{
 \includegraphics[width=1.6\columnwidth]{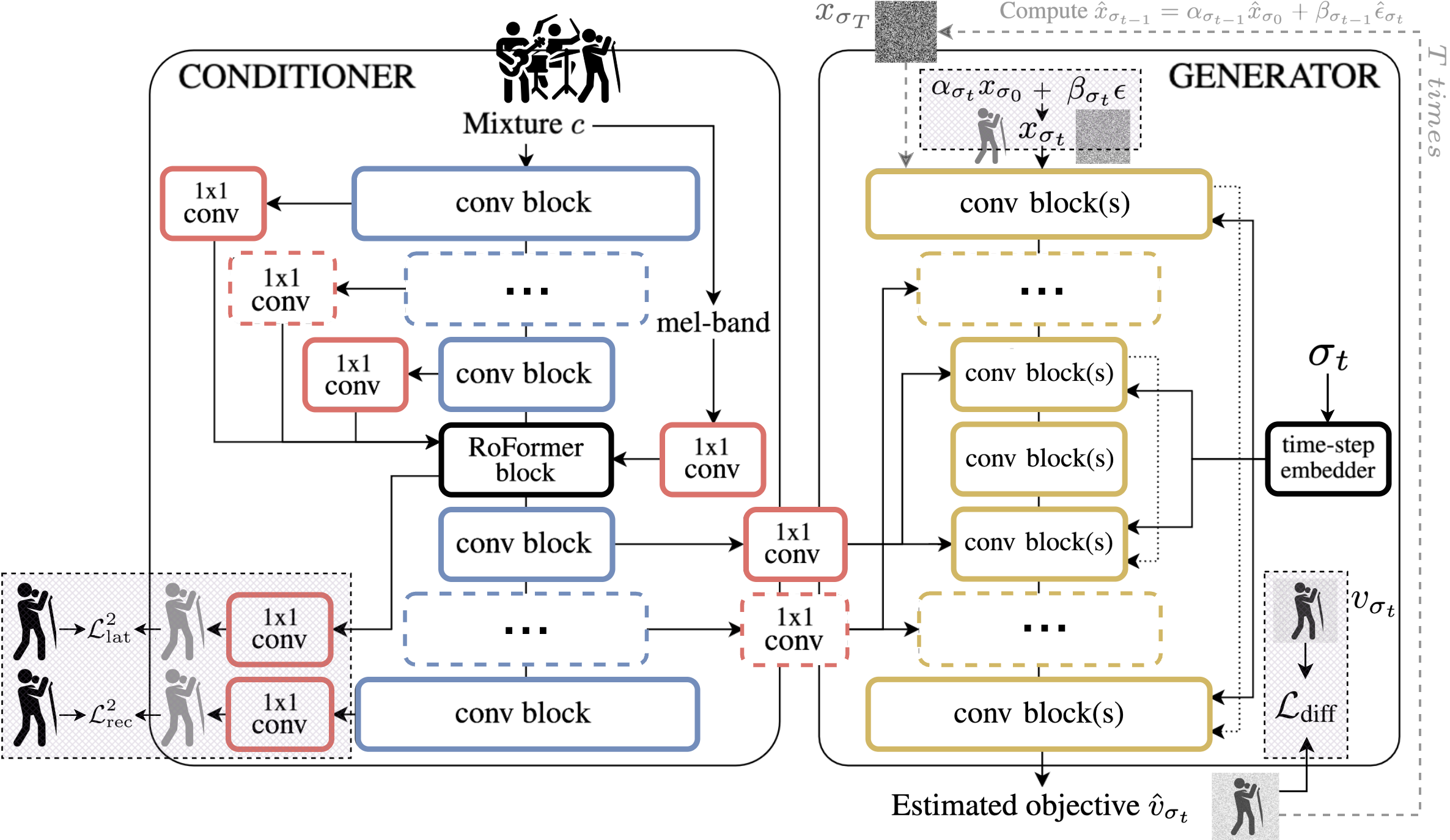}}
 \vspace{-0.2cm}
 \caption{\textbf{Diagram of the system.} The conditioner extracts multi-resolution features from the input mixture $c$. The generator, conditioned on these features, learns the score function to iteratively denoise a Gaussian sample $x_{\sigma_T}$ over $T$ steps, producing the separated vocal signal. Cross-hatched boxes denote components used only during training; thin black dashed lines represent skip connections; the gray dashed line illustrates the inference denoising trajectory.}
 \label{fig:model}
\end{figure*}

\vspace{0.1cm}
\section{Method}\label{sec:method}
\subsection{Diffusion process}\label{sec:diffusion}
\fix{We leverage a diffusion process defined} by a Markov chain of $T$ steps that converts a 2-channel audio signal $x_{\sigma_0} \in \mathbb{R}^{2 \times D} \sim p(x_{\sigma_0})$, \fix{being $p$ a given data distribution from where $x_{\sigma_0}$ is drawn and $D$ audio length in samples}, into a sample of Gaussian noise
when reaching step $T$. This process is controlled \fix{by $\sigma_t$, an instance from a noise schedule of $T$ ordered, ascending values $\in [0, 1]$.}
We adopt $v$-objective diffusion~\cite{salimans2022progressive}, which defines an arbitrary forward diffusion step as:
\begin{equation}
x_{\sigma_t} = \alpha_{\sigma_t}x_{\sigma_0} + \beta_{\sigma_t}\epsilon,
\end{equation}
where $\epsilon \in \mathbb{R}^{2 \times D} \sim \mathcal{N}(0, 1)$ is a random Gaussian sample.
To compute $\alpha_{\sigma_t}$ and $\beta_{\sigma_t}$,
we define $\phi_t := \frac{\pi}{2}\sigma_t$, and obtain the corresponding trigonometric values: $\alpha_{\sigma_t} := \cos(\phi_t)$ and $\beta_{\sigma_t} := \sin(\phi_t)$.

In essence, the forward diffusion process incrementally introduces small amounts of Gaussian noise to $x_{\sigma_0}$ until reaching an isotropic Gaussian sample at step $T$.
Simultaneously, a neural network is trained to \fix{learn to reverse this process}. Intuitively, the network is learning to map points from a Gaussian distribution to an approximated distribution $\hat{p}(x_{\sigma_0})$, from where observations 
can later be sampled. 

Let $v_{\sigma_t} \in \mathbb{R}^{2 \times D}$ be the \textit{velocity} objective, which corresponds to 
an unified prediction target combining noise and data to guide diffusion more effectively.
The objective $v_{\sigma_t}$ is formally computed as:
\begin{equation}
    v_{\sigma_t} = \alpha_{\sigma_t}\epsilon - \beta_{\sigma_t}x_{\sigma_T}
\end{equation}

To optimize the $v$-objective diffusion network $m_\theta$ with trainable parameters $\theta$, we refer to the following loss:
\begin{gather}
\mathcal{L}_\text{diff} = \mathbb{E}_{t\sim[0, T],\sigma_t,x_{\sigma_t}} \big[||m_\theta(x_{\sigma_t}, \sigma_t, c) - v_{\sigma_t} ||_{2}^{2}\big] ,\label{eq:elbo}
\end{gather}
where $\mathbb{E}$ denotes expectation. 
The input signal $c \in \mathbb{R}^{2 \times D}$ is the \fix{\textit{conditioning signal}, which is given to the diffusion model to guide the generation process toward a desired output.}
While MSS has been addressed by unconditionally training a diffusion system and then conditioning the posterior at inference~\cite{mariani2023multi}, we architecturally condition the generation process both at training and sampling, prioritizing quality, isolation, and stronger plausibility with the conditioning mixture given, as input, by the user~\cite{plaja-ldm}.

\subsection{Sampling process}\label{sec:sampling}
The sampling process progressively models a sample \fix{drawn from} approximated distribution $\hat{p}(x_{\sigma_0})$ by denoising a Gaussian noise instance, following the gray dashed line in Figure~\ref{fig:model}.
Prior audio generation works achieved satisfactory balance between sampling steps and generation quality using the Denoising Diffusion Implicit Models (DDIM) sampler~\cite{schneider2023mousai}.
The DDIM sampling process is performed using arbitrary $T$, and initiates at $\sigma_T=1$~\cite{ddim}.
Increasing $T$ provides a more granular sampling, which to some extent results into \fix{higher-quality} generations.
However, architecturally conditioned DDPM for inverse tasks may require small $T$~\cite{universe_serra}. 
%
To run a given sampling step we first run a forward pass of diffusion model $m_\theta$ to predict $\hat{v}_{\sigma_t}$: 
\begin{equation}
\hat{v}_{\sigma_t} = m_{\theta}(x_{\sigma_t}, \sigma_t, c),
\end{equation}
now using $\hat{v}_{\sigma_t}$, we compute the estimated target sample at $t=0$, denoted $\hat{x}_{\sigma_0}$, and $\hat{\epsilon}_{\sigma_t} \in \mathbb{R}^{2 \times D}$, corresponding to the noise at step $t$: 
\begin{gather}
\hat{x}_{\sigma_0} = \alpha_{\sigma_t}x_{\sigma_t} - \beta_{\sigma_t}\hat{v}_{\sigma_t} \label{eq:pred_x} \\ 
\hat{\epsilon}_{\sigma_t} = \beta_{\sigma_t}x_{\sigma_t} + \alpha_{\sigma_t}\hat{v}_{\sigma_t}
\end{gather}

For $t>0$, the input for the next sampling step is defined as:
\begin{equation}\label{eq:next_sampling_step}
\hat{x}_{\sigma_{t-1}} = \alpha_{\sigma_{t-1}}\hat{x}_{\sigma_0} + \beta_{\sigma_{t-1}}\hat{\epsilon}_{\sigma_t},
\end{equation}
and for $t=0$, we keep the output of Equation~\ref{eq:pred_x}.
%
%

Note that this sampling process is \fix{inherently} deterministic. This \fix{enables the generation of high-fidelity outputs in significantly less steps, making the DDIM sampler an ideal suit for our well-defined problem.}
%
\fix{However, we hypothesize} that the rigidness of this iterative modeling process may lead to erroneous or failed generation of parts of the target signal, which may not be recoverable through the inherent noise \fix{that is being predicted and summed back}. 
We propose to introduce stochasticity during sampling–-as formerly done for DDIM sampling for music and image to enrich creativity~\cite{ddim}–-\fix{aiming at providing the system with extra, fresh material to mitigate over-modeling and potentially restore lost signal components.}

To \fix{introduce} the random refinement noise, we first compute the refinement scaling factor $\delta_{t-1}$, and rescale $\beta_{\sigma_{t-1}}$ to prevent the overall noise from being excessively strong for the model to process~\cite{ddim}:
\begin{gather}
    \delta_{\sigma_{t-1}} = \eta \sqrt{\frac{\beta_{\sigma_{t}}^2}{\beta_{\sigma_{t-1}}^2}} \sqrt{1 - \frac{\alpha_{\sigma_{t-1}}^2}{\alpha_{\sigma_{t}}^2}} \\
    \beta'_{\sigma_{t-1}} = \sqrt{\beta_{\sigma_{t}}^2 -  \delta_{\sigma_{t-1}}^2},
\end{gather}
where $\eta$ parametrizes the stochasticity level.
We enable the option to filter the stochastic refinement noise.
\fix{While} one could chose any frequency range, we propose to use a 4$^{th}$-order Butterworth
high-pass filter (HPF) with cutoff frequency $f_c$. Therefore, we attenuate the stochasticity in the low- and mid-frequency range and confine the refinement on the high-frequency end of the spectrum, which we hypothesize to be more challenging to generate, and is where the waveform separation artifacts~\cite{waveunet} and the muffled effect from frequency masking models typically lie~\cite{msg}. 
\fix{Prior work introduced global high-frequency emphasis via purple-noise driving within the diffusion sampling process~\cite{tachibana-solver}. In contrast, we apply controlled filtered stochastic noise at high frequencies targeting a more tailored refinement without modifying the underlying diffusion solver.}

Now, being $\epsilon_{\text{ref}} \in \mathbb{R}^{2\times D}\sim \mathcal{N}(0, 1)$ the refinement noise, we reformulate Equation~\ref{eq:next_sampling_step}:
\begin{equation} \label{eq:next_sampling_step_2}
    \hat{x}_{\sigma_{t-1}} = \alpha_{\sigma_{t-1}}\hat{x}_{\sigma_0} + \beta'_{\sigma_{t-1}}\hat{\epsilon}_{\sigma_t} + \delta_{\sigma_{t-1}}\text{HPF}(\epsilon_{\text{ref}}, f_c),
\end{equation}
here, \fix{$\text{HPF}(\epsilon_{\text{ref}}, f_c)$ is variance-normalized, ensuring its total power aligns with the theoretical requirements of DDIM sampling.}

\subsection{Generation network}\label{sec:generation_network}
To approximate $p(x_{\sigma_0})$ we use a 1D convolutional U-Net, given their established reliability for MSS and diffusion modeling~\cite{diff_beat_gans} and ability to preserve structure~\cite{unet}. It corresponds to the network in the right in Figure~\ref{fig:model}.
Each level of the generator U-Net includes $b$ residual blocks of $2$ convolutional layers.
Each convolutional layer is preceded by group normalization and SiLU activation. After these, self-attention is optionally included to focus on the time-wise relationships, enriching the context.
To inject the diffusion step $\sigma_t$, we compute 1024-channeled random Fourier feature embeddings, which are processed using a 3-layer multi-layer perceptron with GELU activations, and finally injected using FiLM layers~\cite{perez2018film}. The embedded diffusion step is injected after the two convolutional layers.

The last component of a generator block is an $n \times n$ convolutional layer to downsample or upsample the feature vectors in the encoder and decoder respectively. Factor $n$ indicates the compression or expansion ratio. 
In case no down or upsampling is desired for a given block, this layer is inserted with $n=1$. If $n>1$, the feature channels are doubled when downsampling, and divided by 2 otherwise.

\subsection{Conditioner network}\label{sec:conditioner_network}
To process the conditioning signal $c$, we use a coupled network based on the conditioner module in UNIVERSE~\cite{universe_serra},
%
for which we build our version from scratch while proposing modifications motivated by the MSS literature. It corresponds to the left system in Figure~\ref{fig:model}.

\textbf{General structure.} The conditioner is an autoencoder composed of residual convolutional blocks.
In this case, we strictly have a single block per depth level.
These blocks are composed of three convolutional layers, each preceded by a PReLU activation and group normalization, the latter providing a more stable training.
We propose to introduce a time-wise self-attention mechanism to certain depth levels, enabling the model to capture long-range dependencies.
Finally, a convolutional layer is used to compress or expand the time resolution of the feature vectors. The feature channels are converted accordingly following the exact same principle than the generator U-Net (see last paragraph of Sec.~\ref{sec:generation_network}), and always by a factor of 2.
%


\textbf{Skip-connections.} U-Net-type skip connections are not used in the conditioner, aiming at reducing the flow of interferences to the final conditioning signal. However, these are replaced by auxiliary connections that extract the output from each encoder block, process it using a $1\times1$ convolutional layer preceded by group normalization and ReLU activation, and inject the resulting vector into the bottleneck via summation~\cite{universe_serra}. These are crucial for preserving structure and context, which is essential for MSS~\cite{unet}. The feature vectors for the auxiliary connections are retrieved from the output of each encoder block, after the residual connection.

\textbf{Conditioning mechanism.} To inject the condition to the generator, we rely on a multi-resolution conditioning approach~\cite{universe_serra, controlnet}, which assumes that both generator and conditioner are autoencoders that share depth and time compression ratios at all levels. 
We retrieve an embedding from each decoder block in the conditioner (we propose to retrieve it from the block output). We propose to inject it to the corresponding level in both the encoder and decoder of the generator, aiming at further guidance. Auxiliary $1\times1$ convolutions with no padding are used to match the feature channels, which may not necessarily be the same between generator and conditioner in a given level of network depth.
We propose to inject the multi-resolution embeddings by concatenating along the feature channels, followed by a $1\times1$ convolution to dynamically merge the features.

\textbf{Further upgrades.} We propose to enhance the conditioner network by inserting a 6-layer transformer with rotary embeddings~\cite{su2021roformer}, leveraging their effectiveness in modeling time-wise dependencies for MSS even using $<10$h of training data~\cite{bs_roformer}. A transformer has been also introduced in the bottleneck of a U-Net~\cite{ht_demucs}, in this case providing notable improvement when trained with extra data.
We use flash-attention~\cite{dao2022flashattention} to alleviate the increase of computational demands.

\subsection{Auxiliary loss terms}\label{sec:aux_loss}
We incorporate two auxiliary loss terms out of the primary gradient flow, enabling end-to-end training of the entire system while optimizing the conditioner for separation, similarly to~\cite{universe_serra}.
%
Let $C_{\text{enc}}$ and $C_{\text{dec}}$ be the conditioner encoder and decoder respectively. We define a $1 \times 1$ convolutional layer $H_N:\mathbb{R}^{N\times B} \rightarrow \mathbb{R}^{2\times D}$, 
where $N$ are feature channels and $B$ feature size.
We define the first loss term
as:
\begin{equation}
\mathcal{L}^{2}_{\mathrm{lat}} = \frac{1}{D} \sum_{d=1}^{D} 
    \left\lVert  
        x_{\sigma_0 d}\;-\;H_{N}(C_{\mathrm{enc}}(c)_d
    \right\rVert^2
\end{equation}

This loss term is aimed at optimizing the compressed latent \fix{vector} towards the separated vocal by comparing  $\hat{x}_{\text{lat}}\in\mathbb{R}^{2\times D}$ obtained using $H_N$ and the ground-truth vocals \fix{using using L$2$ loss.}
A second loss term denoted $\mathcal{L}^{2}_{\text{rec}}$
is calculated in the decoder output, penalizing again the training process if the embedding contains non-vocal components, and functioning also as autoencoder reconstruction loss:
\begin{equation}
\mathcal{L}^{2}_{\mathrm{rec}} = \frac{1}{D} \sum_{d=1}^{D} 
    \left\lVert  
        x_{\sigma_0 d}\;-\;H_{N}(C_{\text{dec}}(C_{\text{enc}}(c))_d
    \right\rVert^2
\end{equation}


\vspace{0.03cm}
\section{Experiments}\label{sec:experiments}

\subsection{Experimental setup}\label{sec:exp_setup}
Both generator and conditioner have seven depth levels, with downsampling factors $\{1, 2, 4, 1, 4, 1, 4\}$, factor of $1$ indicating no compression or expansion. The generator has a single block in its first three levels, while the remaining levels contain two blocks each. The self-attention mechanism is applied to the four deepest levels of both networks.

We hypothesize that computing an effective multi-resolution embedding is more challenging than learning the score function. Therefore, we allocate greater capacity to the conditioner, using $128$ feature channels at the first level
while the generator network is designed with $32$ input channels. 
The feature channels are doubled (or halved in the decoder) at the network depth levels corresponding to a downsampling factor $> 1$.
With that in mind, the convolutional heads for the auxiliary losses are defined as $H_{2048}$ for $\mathcal{L}^{2}_{\text{lat}}$, and $H_{128}$ for $\mathcal{L}^{2}_{\text{rec}}$.
The multi-resolution conditioning is injected to the four deepest levels of the generator, which are those including time-wise self-attention. 
%


We operate on stereo audio at a rate of 44.1kHz.
Assuming that generating silence is simpler, during training we discard $95\%$ of the silent vocal samples. We use data augmentation: polarity inversion, channel flipping, pitch shifting, and stem remixing.
We use batch size of $32$, AdamW optimizer with a weight decay of $1*10^{-3}$, and cosine annealing scheduler with linear warm-up to stabilize the training. 
We train the system for 1M steps. It takes a week on four A100 GPU. The conditioner module has 82.9M parameters, and the generator 15.9M.
To sample, we process the tracks in $3$s chunks using 20\% overlap-add, where we find an optimal computation vs. quality compromise.

\subsection{Datasets}
DDPM benefit greatly from large training data~\cite{diff_beat_gans}, however, multi-stem datasets are significantly expensive to collect. To evaluate the learning capacity of our system on open data, we train our system using musdb18hq~\cite{musdb}, which includes 100 training tracks ($\approx6$h). Nonetheless, we also experiment with a private collection of $\approx 400$h of multi-stem music including pairs of mixtures and singing vocals, aiming at exploring the full potential of our system.
Related studies on diffusion models for audio have relied on collections notably larger than ours (e.g. $1,500$h~\cite{universe_serra}, $2,500$h~\cite{schneider2023mousai}, or $19,500$h~\cite{evans2024fast}).

\subsection{Evaluation procedure}
\vspace{-0.025cm}
We compute the utterance-level Signal-to-Distortion ratio (SDR) and report the median over the musdb18hq testing set, following practices of prior generative separation work~\cite{instglow,mariani2023multi}.
%
We rely on the implementation from the Sound Demixing Challenge 2023~\cite{sdx_2023}.
To account for the stochasticity in the sampling process if $\eta>0$, we evaluate each model five times and report the mean.

We test three versions of our system:
\textit{Diff-DMX-RT-musdb} (trained using musdb18hq plus self-attention and rotary transformer in the conditioner),
\textit{Diff-DMX-extra} (private collection, no attention nor enhanced bottleneck in the conditioner),
and \textit{Diff-DMX-RT-extra} (private collection plus attention and rotary transformer).
We compare against GAN-prior~\cite{gan_prior}, InstGlow~\cite{instglow}, and MSDM~\cite{mariani2023multi}, all generating waveforms.
We train MSDM for vocal separation \fix{using musdb18hq and following the instructions in the companion repository}, since no weights to separate the singing voice are available online.
%

\begin{table}[t!]
\centering
\caption{\textbf{Median utterance-level SDR on the musdb18hq testing set.} Comparison of our system with generative baselines and the IRM oracle. We provide number of required forward passes during inference ($\mathbf{T}$) and model parameters (\textbf{\#Params}, approximated to the nearest million). All models are trained using musdb18hq, unless indicated in the \textbf{Extra data} column. HT-Demucs~\cite{ht_demucs} serves as the non-generative reference.}
\vspace{0.04cm}
{\fontsize{8}{9.5}\selectfont
\setlength{\tabcolsep}{4.5pt}
\renewcommand{\arraystretch}{0.92}
\begin{tabular}{@{}lcccc|c@{}}
\toprule
\textbf{Model} & \textbf{Gen?} & $\mathbf{T}$ & \textbf{\#Params} & \textbf{Extra data} & \textbf{SDR} (dB) \\
\midrule
IRM oracle~\cite{sisec2018} & -- & \xmark & N/A & N/A & 9.43 \\
\midrule
GAN-prior~\cite{gan_prior} & \checkmark & 1000 & 121M & \xmark & -0.44 \\
InstGlow~\cite{instglow} & \checkmark & 150 & 13M & \xmark & 3.46 \\
MSDM~\cite{mariani2023multi} & \checkmark & 150 & 405M & \xmark & 3.64 \\
\midrule
Diff-DMX-RT-musdb & \checkmark & 50 & 99M & \xmark & 5.38 \\
Diff-DMX-extra & \checkmark & 50 & 71M & $\approx$400h & 6.00 \\
Diff-DMX-RT-extra & \checkmark & 50 & 99M & $\approx$400h & \textbf{8.77} \\
\midrule
HT-Demucs~\cite{ht_demucs} & \xmark & 1 & 346M & \xmark & 7.93 \\
HT-Demucs~\cite{ht_demucs} & \xmark & 1 & 346M & $\approx$53h & 9.37 \\
\bottomrule
\end{tabular}
}

\label{tab:results}
\end{table}

\vspace{0.05cm}
\section{Results} \label{sec:results}
\subsection{Separation performance}
See Table~\ref{tab:results} for the objective separation evaluation and comparison.

\textbf{SDR.} 
Under the same training data, our system outperforms the generative baselines.
Despite the disadvantage of generative models on SDR~\cite{gan_prior,pritish_sep}, our system supplemented with additional training data achieves competitive performance against HT-Demucs, the non-generative reference.
The combination of wider context and the proposed upgrades in the conditioner network provide an important performance boost, objectively showing the generative and separation potential of the proposed system.
The proposed upgrades allow the \textit{Diff-DMX-RT-musdb} model to perform comparably to \textit{Diff-DMX-extra}–-which uses additional training data but does not incorporate the network modifications–-resulting only in a difference of $0.62$dB.

\textbf{Efficiency.} Our model is the second lightest and is the generative system using less sampling steps to reach the performance peak.
Inspired by \cite{mariani2023multi}, we calculate the inference time for a $12$s chunk. In a 12GB GPU, InstGlow needs $16.50\pm0.12$s and MSDM $11.95\pm0.14$s, while our top system using $T=50$ takes $8.45\pm0.03$s, being the fastest among the generative models with off-the-shelf inference code.

\textbf{SIR.} Prior generative separation works report Signal-to-Interference Ratio to study how clean from accompaniment the vocal estimations are~\cite{pritish_sep, gan_prior}. \textit{Diff-DMX-RT-extra} reaches $21.37$dB and levels the IRM oracle, which scores $20.78$dB.
Ultimately, the generator network is trained to model the distribution of solo vocals. The results suggest that this design choice contributes to the achieved level of cleanliness.

\subsection{Sampling process ablation}
See the sampling ablation study in Table~\ref{tab:ablation}.

\textbf{Sampling steps}. As observed for  $T=100$, increasing the number of steps does not necessarily yield better separation quality.
In contrast, using $50$ steps consistently improves over $20$ steps.
%
However, this is not by a large margin, suggesting that architecturally conditioned diffusion together with the DDIM sampler have the potential to model the target in a few steps, while later iterations mainly refine~\cite{universe_serra}.

\textbf{Stoch. refinement noise: strength ($\eta$) and range ($f_c$)}. In terms of $\eta$, the most significant improvement is found at $0.4$, particularly when the stochastic noise is high-pass filtered. While $\eta=0.2$ provides a more moderated refinement, \fix{higher stochasticity (e.g. $\eta = 0.8$) deviates the generation too strongly from the expected output.}

Introducing stochasticity uniformly across the entire frequency range does not yield a positive effect on generation quality. In contrast, we achieve actual improvement when the refinement noise is high-pass filtered. 
We observe that attenuating the stochastic noise only below $600$Hz (where the fundamental frequency of modern melodies is normally found) provides a limited improvement, but we consistently enhance the results by moving the cutoff to $f_c=2$kHz and $5$kHz.
This suggests that \fix{moderately introducing extra stochastic noise} in the high-frequency end–-where the model is potentially less accurate–-can be beneficial.
The high-frequency vocal content is more sparsely distributed and have lower energy, and the model can exploit the added variability to increase the overall fidelity.

\fix{Meanwhile}, attenuating the noise in the low- and mid-frequency end contributes to retain structure where the system shows higher capabilities, \fix{contributing also to cleaner generation overall for smaller $T$ or for cases where the model does underperform.}
Nonetheless, the proposed frequency-selective regularization is not always capable of completely preventing high-frequency artifacts arising particularly in tracks with intrusive accompaniments or uncommon vocal effects. This is a remaining challenge \fix{to address in further work.}

\begin{table}[t]
\centering
\caption{\textbf{Ablation study on the sampling algorithm.} We evaluate the number of sampling steps ($T$), level of stochasticity ($\eta$), and cutoff frequency for the added stochastic noise ($f_c$). We use \textit{Diff-DMX-RT-extra}.}
\vspace{-0.05cm}
\small
\setlength{\tabcolsep}{3pt}
\renewcommand{\arraystretch}{0.95}

\begin{minipage}[t]{0.31\columnwidth}
\centering
\textbf{$T = 20$} \\[2pt]
\vspace{0.05cm}
\begin{tabular}{cc|c}
\toprule
$\eta$ & $f_c$ & \textbf{SDR} (dB) \\
\midrule
\xmark & \xmark & 6.86 \\
0.2    & \xmark & 6.85 \\
0.2    & 600 & 6.97 \\
0.2    & 2k & 7.36 \\
0.2    & 5k     & 7.55 \\
0.4    & 5k     & 8.26 \\
\bottomrule
\end{tabular}
\end{minipage}
\hfill
\begin{minipage}[t]{0.31\columnwidth}
\centering
\textbf{$T = 50$} \\[2pt]
\vspace{0.05cm}
\begin{tabular}{cc|c}
\toprule
$\eta$ & $f_c$ & \textbf{SDR} (dB) \\
\midrule
0.2 & \xmark & 7.17 \\  
0.4 & \xmark & 7.25 \\  
0.4 & 600    & 7.59 \\
0.4 & 2k     & 8.64 \\
\textbf{0.4} & \textbf{5k}     & \textbf{8.77} \\
0.8 & 5k     &  8.33 \\
\bottomrule
\end{tabular}
\end{minipage}
\hfill
\begin{minipage}[t]{0.31\columnwidth}
\centering
\textbf{$T = 100$} \\[2pt]
\vspace{0.05cm}
\begin{tabular}{cc|c}
\toprule
$\eta$ & $f_c$ & \textbf{SDR} (dB) \\
\midrule
0.4 & 5k & 8.52 \\
0.8 & 5k & 7.20 \\
\bottomrule
\end{tabular}
\end{minipage}

\label{tab:ablation}
\end{table}


\section{Conclusions}
Deep generative modeling is transforming music production, primarily through music generation systems.
However, its potential for building assistive tools for music creation and analysis remains less explored.
In this work, we examine how diffusion models can be used to perform vocal source separation directly in the waveform domain, conditioning on mixture audio.
We leverage the condition processing module in UNIVERSE, a diffusion model for speech enhancement, for which we build our alternative from scratch.
We tune the sampling process and perform an ablation study to refine the separations.
Our system outperforms existing generative separation models on open multi-stem music datasets, and achieves competitive performance against non-generative models when trained with supplementary data.
We report strong performance on interference removal.
In contrast to non-generative and mask- or transformation-based separation systems operating on time-frequency representations, the configurable sampling algorithm allows the user to explore different sampling parameters to tune the generated result. Users may prioritize, depending on particular needs, quality or inference time or vice versa, and may also refine the separations.
We publish a code base to perform further research.
%


\clearpage
\bibliographystyle{IEEEtran}
\bibliography{refs25}







\end{document}